\documentclass[12pt,preprint]{aastex}

\slugcomment{}

\shorttitle{Sequential Analysis in Particle Astronomy}
\shortauthors{BenZvi et al.}

\begin{document}

%% LaTeX will automatically break titles if they run longer than
%% one line. However, you may use \\ to force a line break if
%% you desire.

\title{Sequential Analysis Techniques for Correlation Studies \\
       in Particle Astronomy}

\author{S.Y. BenZvi\altaffilmark{1}, B.M. Connolly\altaffilmark{2}, and S. Westerhoff\altaffilmark{1}}
\altaffiltext{1}{University of Wisconsin-Madison, Department of Physics, 1150 University
              Avenue, Madison, WI 53706, USA}

\altaffiltext{2}{University of Pennsylvania, Department of Physics and Astronomy,
              209 South $33^{\mathrm{rd}}$ Street, Philadelphia, PA 19104, USA; brianco@sas.upenn.edu}

%\author{S.Y. BenZvi}
%\affil{Columbia University, Department of Physics and Nevis Laboratories,
%                   538 West $\it 120^{th}$ Street, New York, NY 10027, USA}

%\author{B.M. Connolly}
%\affil{University of Pennsylvania, Department of Physics and Astronomy,
%                209 South $\it 33^{rd}$ Street, Philadelphia, PA 19104, USA}

%\author{S. Westerhoff}
%\affil{University of Wisconsin-Madison, Department of Physics, 1150 University
%              Avenue, Madison, WI 53706, USA}

\begin{abstract}

Searches for statistically significant correlations between arrival directions
of ultra-high energy cosmic rays and classes of astrophysical objects are
common in astroparticle physics.  We present a method to test potential
correlation signals of \textit{a priori} unknown strength and evaluate their
statistical significance sequentially, i.e., after each incoming new event in a
running experiment.  The method can be applied to data taken after the test has
concluded, allowing for further monitoring of the signal significance.  It
adheres to the likelihood principle and rigorously accounts for our ignorance
of the signal strength.

\end{abstract}


% -----------------------------------------------------------------------------
\section{Introduction}\label{sec:intro}
% -----------------------------------------------------------------------------

One of the major goals in astroparticle physics is the identification and the
study of sources of ultra-high energy cosmic rays, defined as cosmic rays with
energies larger than $10^{18}$\,eV.  The discovery of discrete sources would
answer longstanding questions about how and where particles are accelerated to
such energies.  So far, no discrete sources have been positively identified.
One major obstacle for the identification of potential sources is the small
number of detected events.  Until a few years ago, the published world data set
of cosmic rays with energies above $4\,\times 10^{19}$\,eV consisted of little
more than 100 events, mainly recorded with the Akeno Giant Air Shower Array
(AGASA) in Japan between 1984 and 2003~\citep{Takeda:1999sg}, and the High
Resolution Fly's Eye (HiRes) Experiment in Utah between 1997 and
2006~\citep{Abbasi:2004ib}.

Nevertheless, the small data set has been subjected to exhaustive searches for
deviations from isotropy.  These include searches for point sources; searches
for an excess of clustering in the distribution of arrival directions on
various angular scales; and searches for correlations with classes of known
astrophysical objects that were considered likely sites of cosmic ray
acceleration.  Some of these searches resulted in potential signals, but
because of the small size of the data set, the statistical significance could
not be established in a reliable manner.  Consequently, while the discovery of
discrete sources was claimed repeatedly, statistically independent data
routinely failed to support earlier claims.  An example is the search for
correlations of cosmic ray arrival directions with objects of the BL Lac
class~\citep{Tinyakov:2001nr,Gorbunov:2004bs,Abbasi:2005qy}.

With a new generation of large-aperture astroparticle physics detectors like
the Pierre Auger Observatory nearing completion in Malarg\"ue, Argentina and
the Telescope Array detector under construction in Utah, the amount of
ultra-high energy data is now growing at an unprecedented pace.  The Pierre
Auger Observatory, for instance, began scientific data taking in January 2004
and has already accumulated over
$9\times10^3\,\mathrm{km}^{2}\,\mathrm{sr}\,\mathrm{yr}$ of integrated
exposure, more than any previous experiment.

\subsection{Basic Search Techniques in Cosmic Ray Physics}

The fact that previous experiments have failed to find statistically
significant deviations from isotropy in skymaps of ultra-high energy cosmic
rays can be seen as an indication that the sources are weak.  In this case,
the most promising correlation searches are not those which aim at finding
sources individually, but rather those conducted on a statistical basis;
i.e., searches for significant correlations of cosmic ray arrival
directions with catalogs of astrophysical objects.   

When studying correlations with objects from a source catalog, one tests
whether the probability $p$ of a given event to arrive from the direction of an
object in the catalog is significantly larger than the probability $p_0$ of the
correlation occurring by chance.  These analyses are typically binned, so an
event is said to correlate with an object from the catalog if the angle between
its arrival direction and the object's position is smaller than some angle
$\theta$.  If the particles are neutral, $\theta$ could be chosen to reflect
the point spread function of the detector.  In the case of cosmic rays,
however, the particles are most likely charged and therefore deflected by
Galactic and intergalactic magnetic fields of (unknown) strength.
Consequently, $\theta$ is usually chosen to be larger than the resolution of
the detector to account for magnetic smearing.

Typically, potential signals are identified after intensive searches using
different angular scales, different energy thresholds, different source
catalogs, and other parameters that are found to maximize the signal strength.
Therefore, an unbiased chance probability for the observed signal can only be
established by discarding the data set used to find the signal and testing the
signal with statistically independent data.  For the test, the source catalog
and all analysis parameters are fixed {\it a priori} to obtain an unbiased
chance probability for the signal.  

Once the \textsl{a priori} analysis parameters are identified, the problem is
easily formulated in terms of a classical hypothesis test, in which new data
are checked for compatibility with a null hypothesis $\mathcal{H}_0$ (``the
data exhibit no significant correlation'') or an alternative ``signal''
hypothesis $\mathcal{H}_1$.  There are several ways to perform such a test.
For example, one can run the test after the new data set has reached a certain
size $n$, or after the experiment has run for a certain fixed amount of time.

Formally, the size of the data set and the acceptance or rejection of the null
hypothesis are determined by two probabilities, $\alpha$ and $\beta$, which are
usually chosen before the start of the test.  These values define the
experimenter's tolerance for different sorts of experimental errors: $\alpha$
is the probability of wrongly rejecting the null hypothesis when
$\mathcal{H}_0$ is true (a type-1 or ``false positive'' error); and $\beta$ is
the probability of wrongly accepting the null hypothesis when $\mathcal{H}_0$
is false (a type-2 or ``false negative'' error).  In a classical one-sided
hypothesis test, where a p-value $P$ is used to estimate the agreement of the
data with the null hypothesis, the result $P<\alpha$ implies rejection of
$\mathcal{H}_0$ at the ``confidence level'' $1-\alpha$.  Meanwhile, the desired
probability of rejecting a false null hypothesis $(1-\beta)$ fixes the required
size of the data set $(n)$.

\subsection{One-Shot vs. Sequential Testing}

If one chooses to evaluate $P$ after a predefined number of events has been
recorded, or a predefined amount of time has elapsed, then the significance of
the signal is tested only once.  However, it is often desirable to evaluate and
test the signal sequentially, i.e., after each new event, rather than
at the end of the test.  This approach allows for the possibility of claiming a
statistically significant result earlier than with methods that check the
signal only once, a distinct advantage when event rates are quite low.  It also
avoids another practical disadvantage of hypothesis tests that arises when the
experiment, for one reason or another, has to discontinue data taking before
the predefined number of events is taken.  In that case, the ``one-shot''
analysis does not lead to a conclusion.

A sequential analysis can be performed in several ways.  If $P$ is evaluated
after every incoming event and not just once after all $n$ events are
collected, a ``penalty'' factor has to be inserted to account for the fact that
there are now more opportunities to satisfy the test by
chance~\citep{Anscombe:1954,Armitage:1969}.  This penalty
factor can be evaluated with simulations and will depend on $n$.  The
dependence of $P$ on $n$ is an undesirable feature of the method; rather than
depending on the data that were actually recorded, $P$ now depends on the
number of events that an observer would have recorded had he decided to perform
a ``one-shot'' test.  The interpretation of the data therefore depends on data
not actually taken.  This feature of the test violates the likelihood
principle~\citep{Berry:1987}.   

In addition, the inclusion of the penalty factor means that data arriving after
the test has ended cannot be used to calculate $P$ for the entire data 
set.  It is therefore not possible to include new data in the calculation
of the probability.  In many practical situations, data taking continues after 
the test has ended, and it is highly desirable to monitor the signal 
probability with new data.  

The classical sequential likelihood ratio test developed by~\citet{Wald:1945,Wald:1947} 
avoids the limitations that arise when using the p-value $P$.  Wald defines the 
likelihood ratio evaluated after the $n^{th}$ event as
\begin{equation}
\mathcal{R}_n=\frac{P(\mathcal{D}|\mathcal{H}_1)}{P(\mathcal{D}|\mathcal{H}_0)}~~,
\label{eq:likelihood_ratio}
\end{equation}
where the denominator and numerator represent the probability of observing a
data set $\mathcal{D}$ given a null hypothesis (no correlation) and an
alternative (correlation).  The ratio $\mathcal{R}_n$ can be evaluated after each
incoming event (i.e. after the $n^{th}$ event) without statistical penalty, and the test stops with the
acceptance or rejection of the null hypothesis when $\mathcal{R}_n$ falls below
or exceeds a predefined value (details will be given in
Section~\ref{sec:method}). Moreover, the evaluation of $\mathcal{R}_n$ can
continue after the decision to see whether new data continue to favor or
disfavor the selected hypothesis. 

The probabilities $P(\mathcal{D}|\mathcal{H}_0)$ and
$P(\mathcal{D}|\mathcal{H}_1)$ in eq.\,(\ref{eq:likelihood_ratio}) depend on the
expected correlations in case of random coincidences and true signals,
respectively.  In correlation studies, the strength of the signal is typically
not known before the test is complete; so in the analysis proposed
by~\citet{Wald:1945,Wald:1947}, one simply takes a ``best guess'' at the lower
bound of the signal strength.  In this paper, we extend Wald's technique to
marginalize the signal strength, which more rigorously accounts for our
ignorance of the true signal.  As in the classical likelihood ratio test, this
extended test can be applied after each new event without statistical penalty,
so that it adheres to the likelihood principle.  It also allows for the
evaluation of the significance of the signal after the test has been fulfilled,
as well as in cases where the test stops prematurely.

We note that the usefulness of this test is not limited to cosmic ray physics.
It can be applied in many other areas of astroparticle physics or astrophysics
where event rates are low, for example in searches for discrete sources of high
energy neutrinos or $\gamma$-rays.

%This paper is organized as follows.  After a description of the method in
%Section\,\ref{sec:method}, we analyze the behavior of the test with simulated
%data sets in Section\,\ref{sec:test}.  Section~\ref{sec:summary} summarizes
%the results.

% -----------------------------------------------------------------------------
\section{The Method}\label{sec:method}
% -----------------------------------------------------------------------------

We consider the case of an analysis searching for correlations between cosmic
ray arrival directions and objects from a catalog.  The background probability
$p_0$ is the probability that a given event correlates by chance.  We want to
test the signal probability $p_1$ against $p_0$.  If two point hypotheses are
tested against each other, $p_0$ and $p_1$ are single numbers; but in general,
$p_1$ can also have a range of values.  If, for example, the ``signal''
corresponds to a stronger correlation than can be expected by chance, then
$p_1>p_0$.

Since an event can either be correlated with an object from the catalog or not,
the probability of observing a data set $\mathcal{D}$ in which $k$ out of $n$
events correlate with sources is given by the binomial distribution
\begin{equation}
P(\mathcal{D}|p) = P(n,k|p) = {n \choose k}\ p^k\ (1-p)^{n-k}
\label{eq:binomial_distribution}
\end{equation}
where $p$ is the probability of a given event to correlate.  If the data show
no significant correlations in addition to those occurring by chance, then
$p=p_0$.

In a sequential analysis that tests hypothesis $\mathcal{H}_1$ against
$\mathcal{H}_0$ with data $\mathcal{D}$, the probability ratio $\mathcal{R}_n$ of
eq.\,(\ref{eq:likelihood_ratio}) is calculated after each incoming event, and is
then compared to two positive constants $A$ and $B$ (where $B<A$).  During each
step in the sequence, the experimenter is presented with the following possible
outcomes:
\begin{enumerate}
  \item $\mathcal{R}_n\ge A$: the test terminates with the rejection of
        $\mathcal{H}_0$.
  \item $\mathcal{R}_n\le B$: the test terminates with the acceptance of
        $\mathcal{H}_0$.
  \item $B<\mathcal{R}_n<A$: the test continues to record data.
\end{enumerate}
\citet{Wald:1945,Wald:1947} showed that the constants
$A$ and $B$ are closely related to the probabilities $\alpha$ and $\beta$ of
type-1 and type-2 errors:
\begin{equation}
  A\leq\frac{1-\beta}{\alpha}~~~\mathrm{and}~~~
  B\geq\frac{\beta}{1-\alpha}~~.
\end{equation}
While it is difficult in most practical situations to estimate exact values for
$A$ and $B$, Wald showed that simply choosing 
\begin{equation}
  A = \frac{1-\beta}{\alpha}~~~\mathrm{and}~~~
  B = \frac{\beta}{1-\alpha}~~, 
\end{equation}
as the test boundaries leads to adequate results if $\alpha$ and $\beta$ are
small (typically, they are not larger than 0.05).  By adequate, we mean that
the true type-1 and type-2 rates will never exceed $\alpha$ and
$\beta$.  In fact, the true error rates will often be smaller than the nominal
$\alpha$ and $\beta$ specified before the start of the experiment.

%The test can terminate at any time and still provide valuable information, 
%or it can continue even after a decision is made to see whether additional 
%data further supports the decision or not.  No penalty factor is required, 
%but still, the probabilities are evaluated after each incoming event.

For a data set that contains $n$ events and $k$ correlations, the likelihood
ratio is given by
\begin{equation}
  \mathcal{R}^\prime _n
    =\frac{P(\mathcal{D}|p_1)}{P(\mathcal{D}|p_0)}
    =\frac{p_1^k (1-p_1)^{n-k}}{p_0^k (1-p_0)^{n-k}}~~.
  \label{eq:likeli}
\end{equation}

In practice, the signal strength $p_1$ is often not known.  We consider here
the common case of a one-sided test where $p_0 < p_1 \leq 1$.  The confidence
in rejecting $\mathcal{H}_0$ typically increases with increasing $p$.  To
evaluate $\mathcal{R}_n$ in this case, we can expand the numerator and
denominator of eq.\,(\ref{eq:likelihood_ratio}) in terms of $p$:
\begin{equation}
  \mathcal{R}_n = \frac{\int_0^1 P(D|p)\ P(p|\mathcal{H}_1)\ dp}
                     {\int_0^1 P(D|p)\ P(p|\mathcal{H}_0)\ dp}~.
\end{equation}

The quantities $P(p|\mathcal{H}_1)$ and $P(p|\mathcal{H}_0)$ represent our
prior assumptions about $p$ in the cases of true signal vs. chance
correlations.  In cosmic ray studies, the probability $p_0$ of a chance
correlation with a catalog object is estimated from the \textsl{a priori}
parameters of the test: e.g., the detector exposure to the catalog,
the angular bin size $\theta$, etc.  In contrast, it is fairly uncommon to have
a reliable estimate of the signal probability $p_1$ beyond the fact that
$p_1>p_0$.  Absent further knowledge of the signal, we can therefore treat the
probability as uniformly distributed on the interval $[p_1,1]$.  Hence, we
summarize our prior knowledge of the two cases by
\begin{eqnarray}
  P(p|\mathcal{H}_1) & = & \frac{\Theta(p-p_1)}{1-p_1}~~, \\
  P(p|\mathcal{H}_0) & = & \delta(p-p_0)~~.
\end{eqnarray}
Note that $p$ is not time-dependent, although we do not see 
anything inherently problematic in inserting a time-dependence.  Although not
many ultra-high energy cosmic ray models propose a time-dependence, 
if a time-dependent model is inserted for $\mathcal{H}_0$, the probability 
of each sucessive event is evaluated based on what is expected 
at the time it was measured.  However, if $\mathcal{H}_0$ 
and $\mathcal{H}_1$ are simply wrong - that is, the
hypotheses do not properly reflect what could happen in nature
- then any result is possible.  This hazard exists for any hypothesis test. 

Solving for the likelihood ratio $\mathcal{R}_n$, we have
\begin{eqnarray}
  \mathcal{R}_n & = & \frac{\int_{p_1}^1 p^k\ (1-p)^{n-k}\ dp}
                         {p_0^k\ (1-p_0)^{n-k}\ (1-p_1)}\\
              & = & \frac{\mathrm{B}(k+1, n-k+1) - 
                          \mathrm{B}(p_1; k+1, n-k+1)}
                         {p_0^k\ (1-p_0)^{n-k}\ (1-p_1)}~,
  \label{eq:final_ratio}
\end{eqnarray}
where $\mathrm{B}(a,b)$ and $\mathrm{B}(x;a,b)$ are the complete and incomplete
beta functions.  Note that eq.\,(\ref{eq:final_ratio}) is a convenient form for
the numerical computation of $\mathcal{R}_n$.

\begin{figure}
\plotone{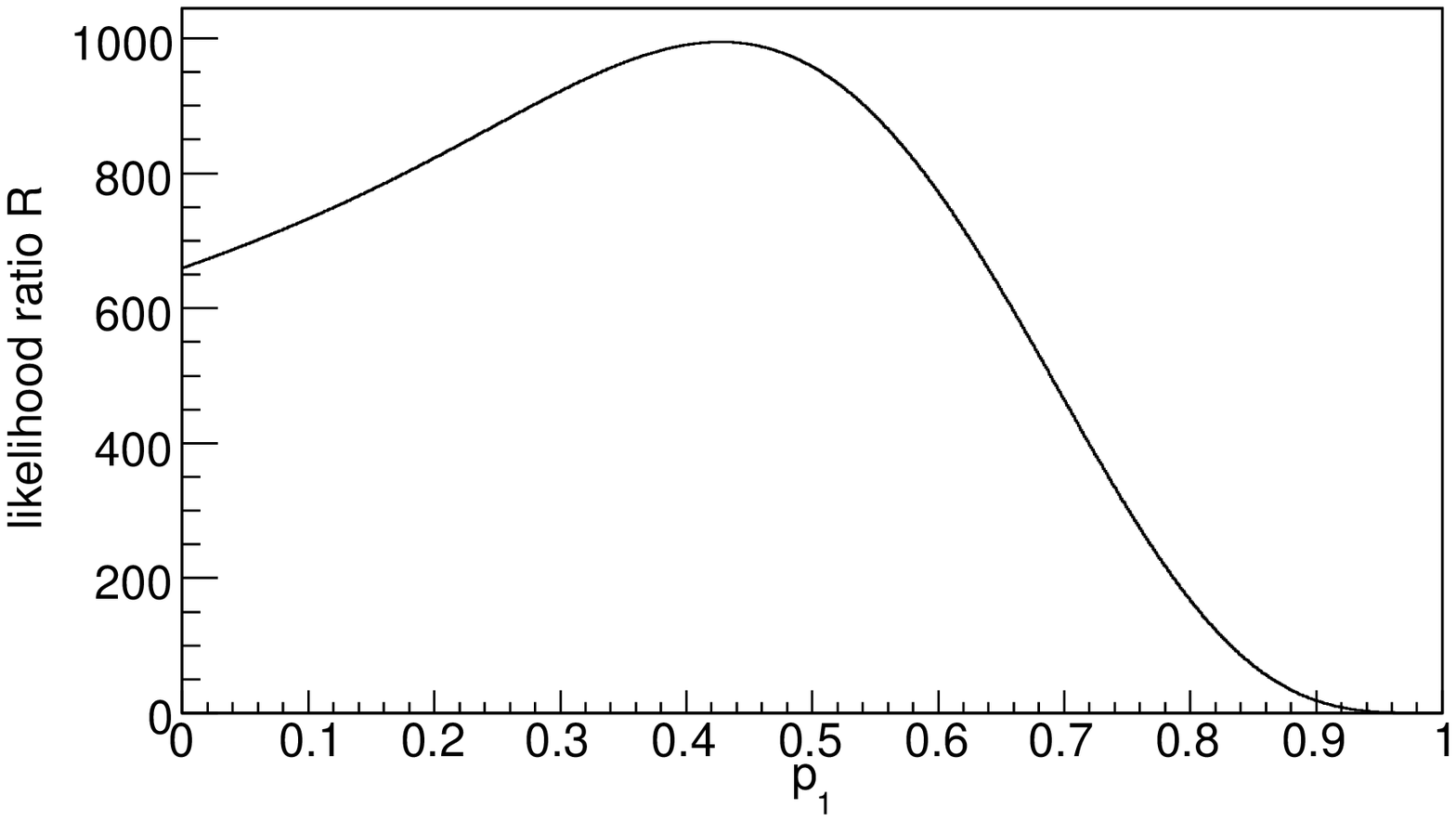}
\caption{Likelihood ratio as a function of $p_1$ for $n=10$, $k=6$, 
and $p_0=0.1$.\label{fig:R_vs_p1}}
\end{figure}

When nothing is known {\it a priori} about the strength of the signal, $p_1$
will be chosen close to $p_0$ to test as large a signal space $p$ as possible.
If more information on $p$ were available --- for example, if it were known
that $p$ is larger than some value $p_{\mbox{\scriptsize min}}$ --- then the range of
integration could be made smaller.  To illustrate the merits of improved
knowledge, Fig.\,\ref{fig:R_vs_p1} shows $\mathcal{R}_n$ as a function of $p_1$
for $n=10$, $k=6$, and $p_0=0.1$.  Since the ``true'' probability for an event
to correlate is $p=6/10=0.6$, choosing $p_1$ close to $p$ increases
$\mathcal{R}_n$ and therefore minimizes the time necessary to confirm the signal.
As $p_1$ continues to increase beyond the true signal probability,
$\mathcal{R}_n$ decreases, as expected.

\begin{figure}
\plotone{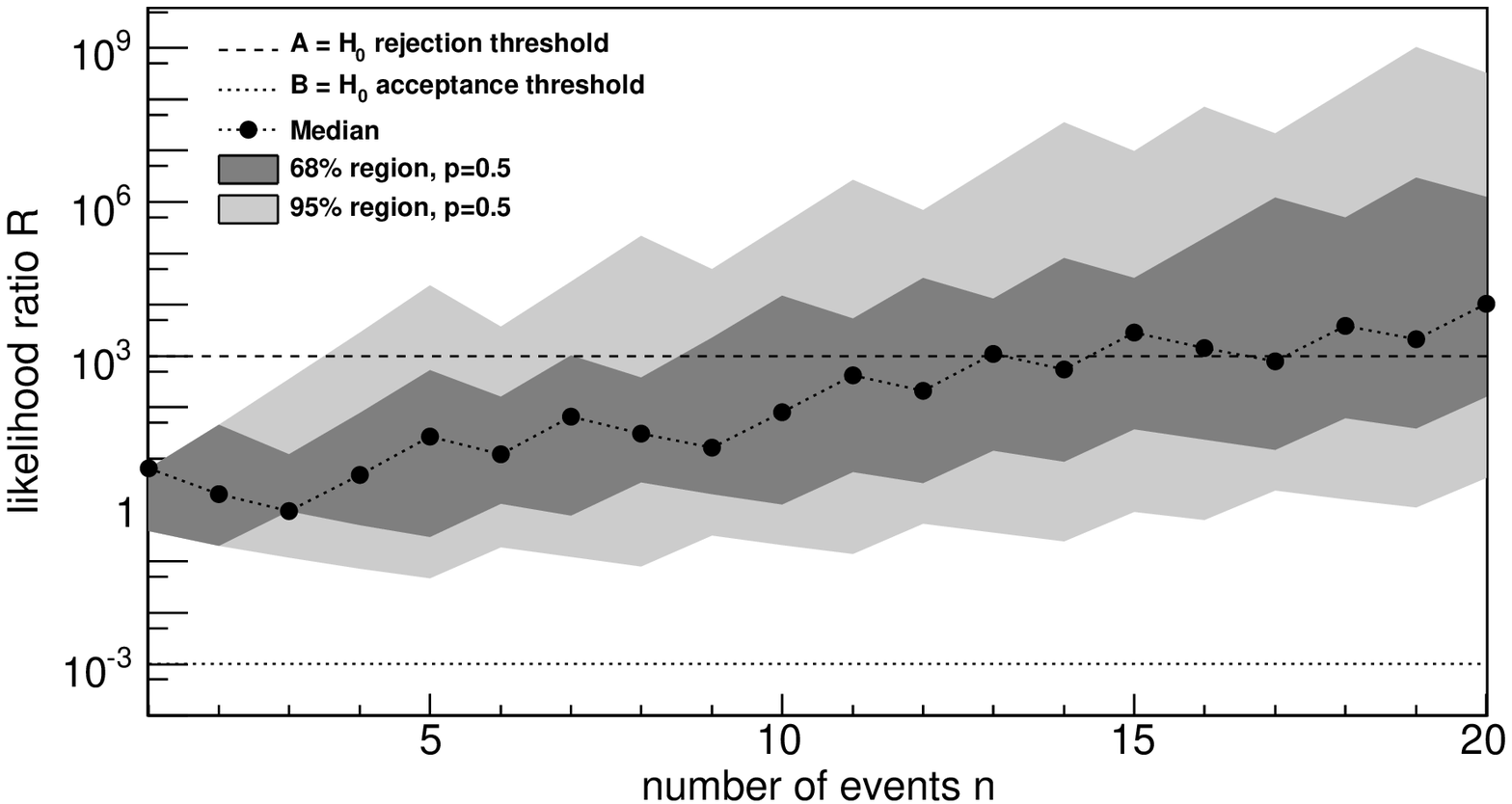}
\plotone{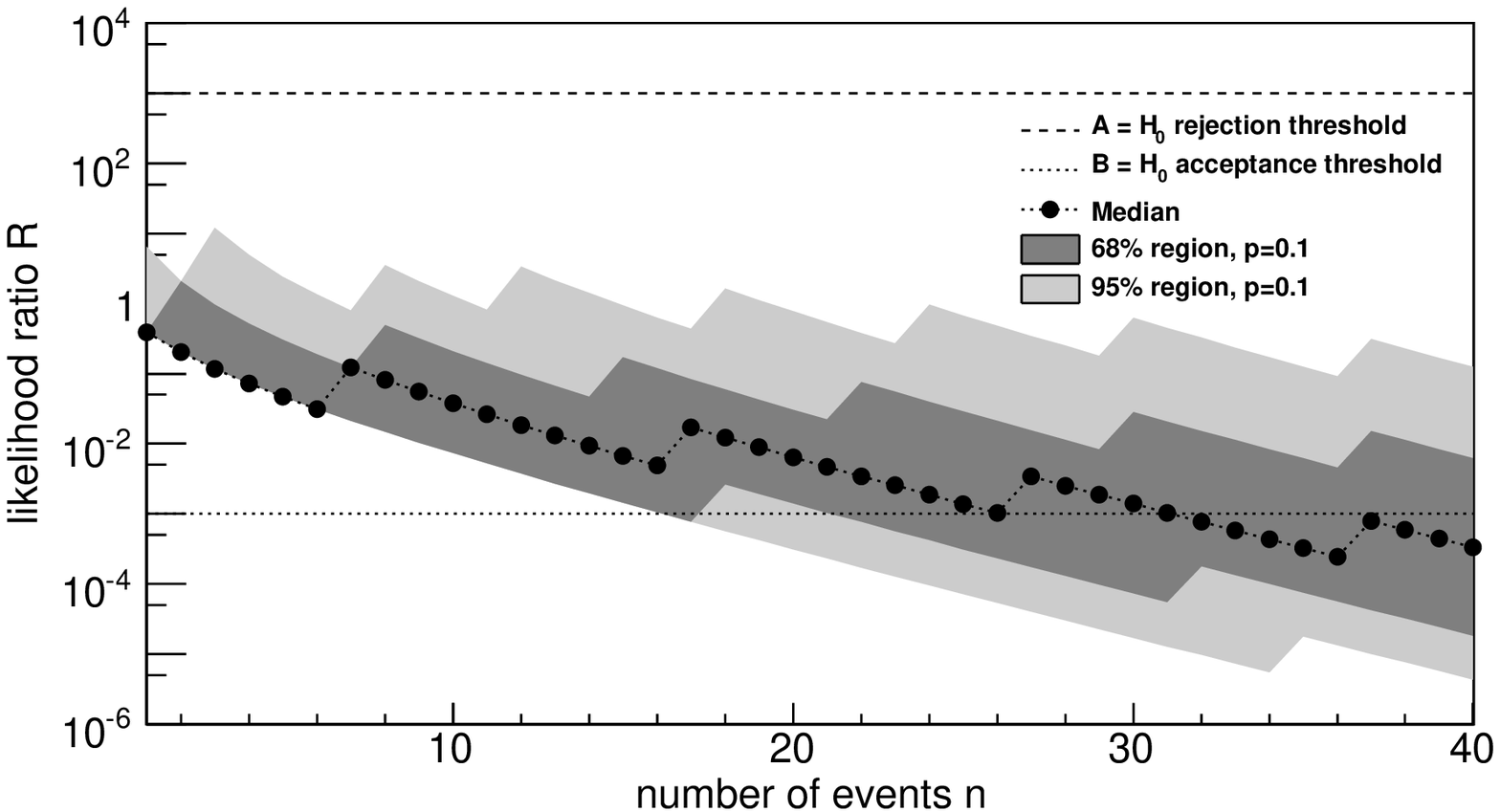}
\caption{Likelihood ratio as a function of the number of events for a background 
probability $p_0=0.1$, $p_1=0.3$, and a signal probability $p=0.5$ (top) and $p=0.1$ (bottom). 
The ratio is calculated for $10^5$ random data sets.  The plots show the median (dark grey dots)
together with the range that includes 68\,\% and 95\,\% of the data sets (dark and light grey
areas).  The values for the test boundaries $A$ and $B$ for $\alpha=\beta=0.001$ are indicated 
as dashed and dotted lines.\label{fig:R_vs_n}}
\end{figure}

Fig.\,\ref{fig:R_vs_n} shows the results of the sequential analysis described
above when applied to simulated data sets.  The background probability is
$p_0=0.1$; $p_1=0.3$ is the minimum signal we choose to distinguish from the
background; and $\alpha=\beta=0.001$.  The upper plot shows the result of the
test for data sets with a correlation probability of $p=0.5$ ($\mathcal{H}_0$
is false), whereas for the bottom plot, $p=0.1$ ($\mathcal{H}_0$ is true).  For
both plots, the analysis is performed for $10^5$ Monte Carlo data sets, and the
dark and light grey areas indicate the range that includes 68\% and 95\% of the
data sets.  

% -----------------------------------------------------------------------------
\section{The Ratio of Likelihoods, the Ratio of Posteriors, and the Meaning
of $\alpha$ and $\beta$}
% -----------------------------------------------------------------------------

Here, $\mathcal{R}_n$ is defined as a ratio of likelihoods, but
one could just as easily define $\mathcal{R}_n$ as a ratio of 
posterior probabilities as suggested by~\citet{Wald:1945,Wald:1947}.  
However, changing the definition
of $\mathcal{R}_n$ carries consequences in the interpretation 
of $\alpha$ and $\beta$.  To understand how, we first review 
what $\alpha$ and $\beta$ mean in the context of the likelihood ratio.

The meaning of the probabilities in the numerator and denominator of 
$\mathcal{R}_n$ are obviously connected to the meaning of $\alpha$ and $\beta$.  
One could argue that, since we are marginalizing parameters anyway, 
we might as well calculate the posterior probabilities as suggested in 
Wald's original paper~\citep{Wald:1945}.  
This has certain advantages.
For instance, the ratio would be defined as 
\begin{eqnarray}
\mathcal{R}^{post}_n = \frac{P(\mathcal{H}_1|D)}{P(\mathcal{H}_0|D)} 
= \frac{P(D|\mathcal{H}_1)P(\mathcal{H}_1)}{P(D|\mathcal{H}_0)P(\mathcal{H}_0)}.
\end{eqnarray}
One could choose priors for $P(\mathcal{H}_1)$ and $P(\mathcal{H}_0)$.
$A$ and $B$ then become thresholds for ``degrees of belief'' that
we must hold for one hypothesis over another before we claim one or the 
other to be true.  
For instance, given that $\mathcal{H}_1$ is true,
$1-\beta$
becomes the required confidence for $P(\mathcal{H}_1|D)$ 
and $\alpha$ the required confidence
for $P(\mathcal{H}_0|D)$ to claim that $\mathcal{H}_1$ is true - i.e.
$A=(1-\beta)/\alpha$.   

However, as noted by~\citet{Wald:1945,Wald:1947}, the likelihood ratio also 
has its merits.  First, the likelihood ratio has some precedent.  Even those 
who subscribe to the Bayesian formalism use marginalized likelihood ratios 
(i.e. Bayes Factors)~\citep{Jeffreys:1939,Kass:1995}; using a likelihood 
ratio avoids the use of priors $P(\mathcal{H}_0)$ and $P(\mathcal{H}_1)$ which 
can strongly influence the result.  Further, likelhood ratios provide 
like comparisons with likelihood ratios used in other analyses with fixed $p_0$ 
and $p_1$.  However, the definitions of $A$ and $B$ become cumbersome even
in the circumstance here where we
are unconcerned whether or not the test ever terminates, 
For instance, given that $\mathcal{H}_1$ is true, 
$A$ parameterizes how much more likely the data must come from a universe
where $\mathcal{H}_1$ is true as opposed to $\mathcal{H}_0$ before 
we claim that $\mathcal{H}_1$ is indeed true. 

In short, using a ratio of posteriors allows 
$\alpha$ and 
$\beta$ to be conceptualized intuitively as degrees of belief
in one hypothesis or another.  Using likelihood ratios is common and, while one
does not have to contend with defining priors for $\mathcal{H}_1$ and $\mathcal{H}_0$,
$\alpha$ and $\beta$ can no longer be conceptualized in terms of degrees of belief
for $\mathcal{H}_0$ and $\mathcal{H}_1$.  
Here, we opt for the more traditional calculation of the likelihood ratio
or what could be thought of as a ratio of posteriors 
if $P(\mathcal{H}_1)=P(\mathcal{H}_0)$.

% -----------------------------------------------------------------------------
\section{Testing the Method}\label{sec:test}
% -----------------------------------------------------------------------------

\subsection{Test Convergence and the Error Rates $\alpha$ and $\beta$}

To account for our ignorance of the true correlation probability $p$ of the
given data set, $p$ is marginalized in the likelihoods in eq.\,(\ref{eq:likeli}).
As mentioned in the previous section, we assume that the signal
probability $p$ that we want to test against the null hypothesis is uniformly
distributed on $\left[p_1,1\right]$.  With no prior knowledge of the signal
other than $p>p_0$, we choose $p_1=p_0$.  

In practice, this approach has an important consequence if one were to
interpret the results of the hypothesis test in terms of the probabilities
$\alpha$ and $\beta$, for example by using $(1-\alpha)$ as a confidence
level for the rejection of the null hypothesis.  Since the numerator now
allows for $p_1<p<1$, $\alpha$ and $\beta$ have, strictly speaking, only
meaning for a data set that has similar properties, i.e. has a correlation
probability that is not a single value, but spread over the interval
$\left[p_1,1\right]$.  However, in reality, any given data set has some fixed
probability $p$ to correlate with objects of a catalog.

Therefore, we must test whether in the case of a fixed $p$ the method returns
probabilities for type-1 and type-2 errors lower than $\alpha$ and $\beta$.  In
general, we expect the type-2 error to be smaller than $\beta$ if the
correlation probability in the data is larger than some minimum value
$p_{\mbox{\scriptsize min}}$.  

A second practical issue is the convergence of the sequential likelihood ratio
test to a conclusion in favor of $\mathcal{H}_0$ or $\mathcal{H}_1$.  When
$p_1=p_0$ and the null hypothesis is true $(p=p_0)$, the ratio test will often
fail to reach a conclusion even as the number of events $n$ becomes quite
large.  
This problem can be avoided in two ways.  One would be to terminate the test after
accumulating some number of events, $n_0$.  The acceptance or rejection of
$\mathcal{H}_0$ would then depend on whether $\mathcal{R}_n$ was greater or less than 1.
However, making a decision in this way would require a modification of 
the type-1 and type-2 errors (see Appendix\,A).  
Another would be to choose $p_1=p_0+\delta$, where
$\delta$ is a positive constant.  
The particular choice of $\delta$ is somewhat
\textsl{ad hoc}, since it mainly reflects the experimenter's degree of belief
about the strength of the signal.  However, for those uncomfortable with this
kind of inference, we present a simple procedure to find $\delta$ such that:
the likelihood ratio $\mathcal{R}_n$ converges to a conclusion while still
satisfying a large number of signal hypotheses; and the type-1 and
type-2 rates of the sequential analysis are consistent with the
classical interpretations of the probabilities $\alpha$ and $\beta$.

%Performing the likelihood ratio test as described above with $p_1=p_0$ leads to
%another practical problem.  In cases where the null hypothesis is true, the
%ratio test often does not come to a conclusion even for large numbers of events
%$n$.  This problem can be avoided by choosing $p_1$ to be larger than $p_0$ by
%some amount $\delta$.  One could go even further by choosing a $\delta$ such
%that the method would not only finish with a finite data set $n$, but with
%type-1 error probabilities smaller than $\alpha$ in data sets where $p$ was
%fixed.  This latter requirement for $\delta$ can be viewed as superfluous for
%two reasons.  First, strictly speaking, $\delta$ is not a parameter that is
%usually found, but rather chosen {\it a priori}; ideally, it should be governed
%by nothing more than the experimenter's degree of belief.  Second, as will be
%discussed below, we could simply pick a $p_1$ above which $\mathcal{R}_n$ has the
%correct $\alpha$ and $\beta$ when the signal probability is larger than $p$.
%This removes the need for scanning over $\delta$ to find an $\alpha$ and
%$\beta$ that behave in the desired fashion.  However, here we discuss a
%procedure to find $\delta$ to satisfy those who seek the best of both worlds: a
%likelihood ratio that leaves the option for a number of signal hypothesis and a
%sequential analysis method that returns an $\alpha$ and $\beta$ that can be
%interpreted intuitively.

In this section, we test these expectations with simulated data sets and
determine values for $\delta$ and $p_{\mbox{\scriptsize min}}$ for some typical values for
$p_0$, $\alpha$, and $\beta$.  If we find $\delta$ to be small and
$p_{\mbox{\scriptsize min}}$ to be close to $p_0$, then the test will terminate with type-1
and type-2 error rates that are smaller than $\alpha$ and $\beta$, giving the
result an intuitive interpretation.  For each of the following tests, we
produce $10^5$ simulated data sets\footnote{We will use $\alpha=\beta=0.001$, and therefore
test the method on $10\times 1/0.001$.} with a correlation probability $p$ and
subject these data sets to a sequential analysis with predefined values for
$\alpha$ and $\beta$.  

\begin{figure}
\plotone{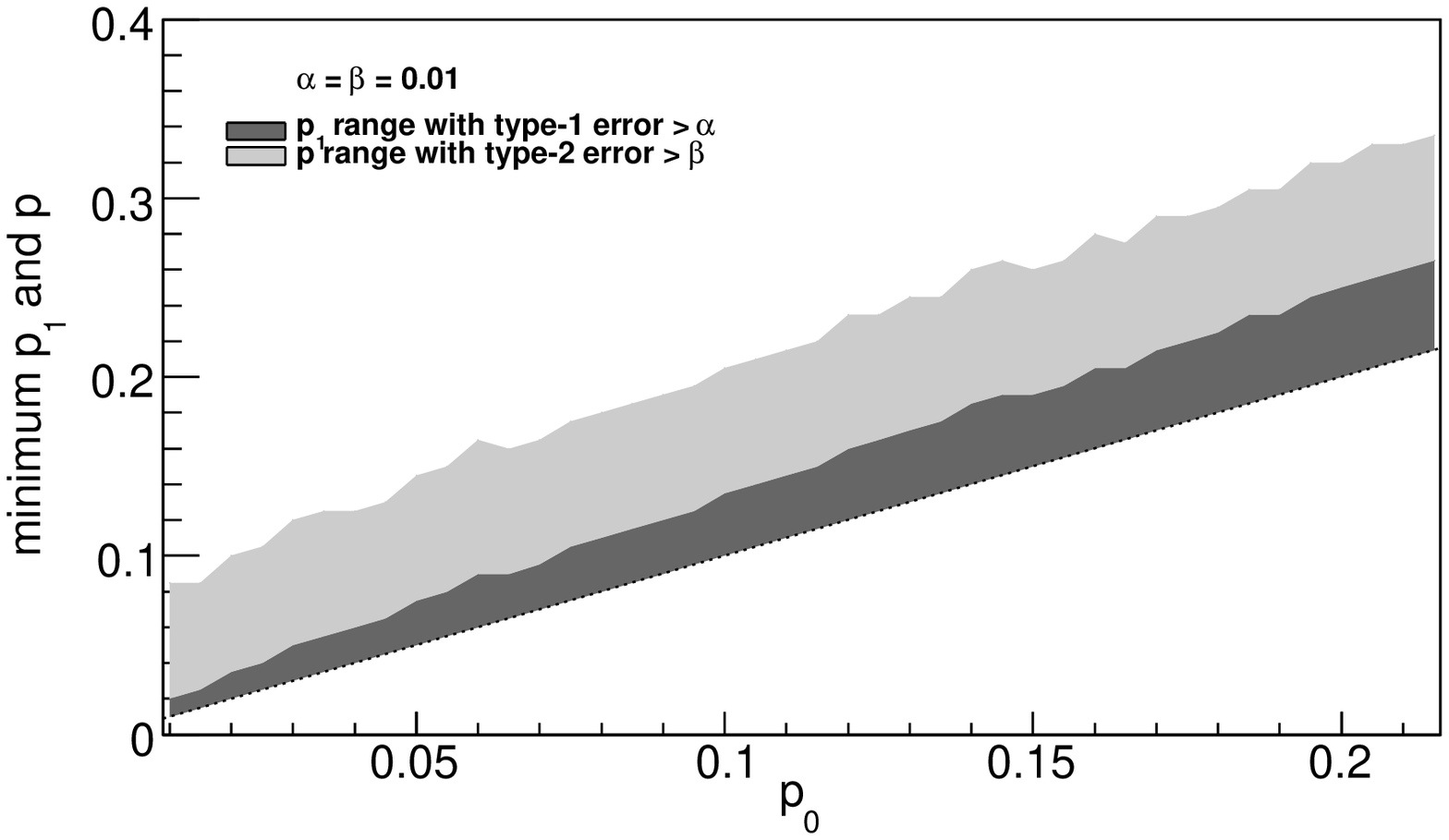}
\plotone{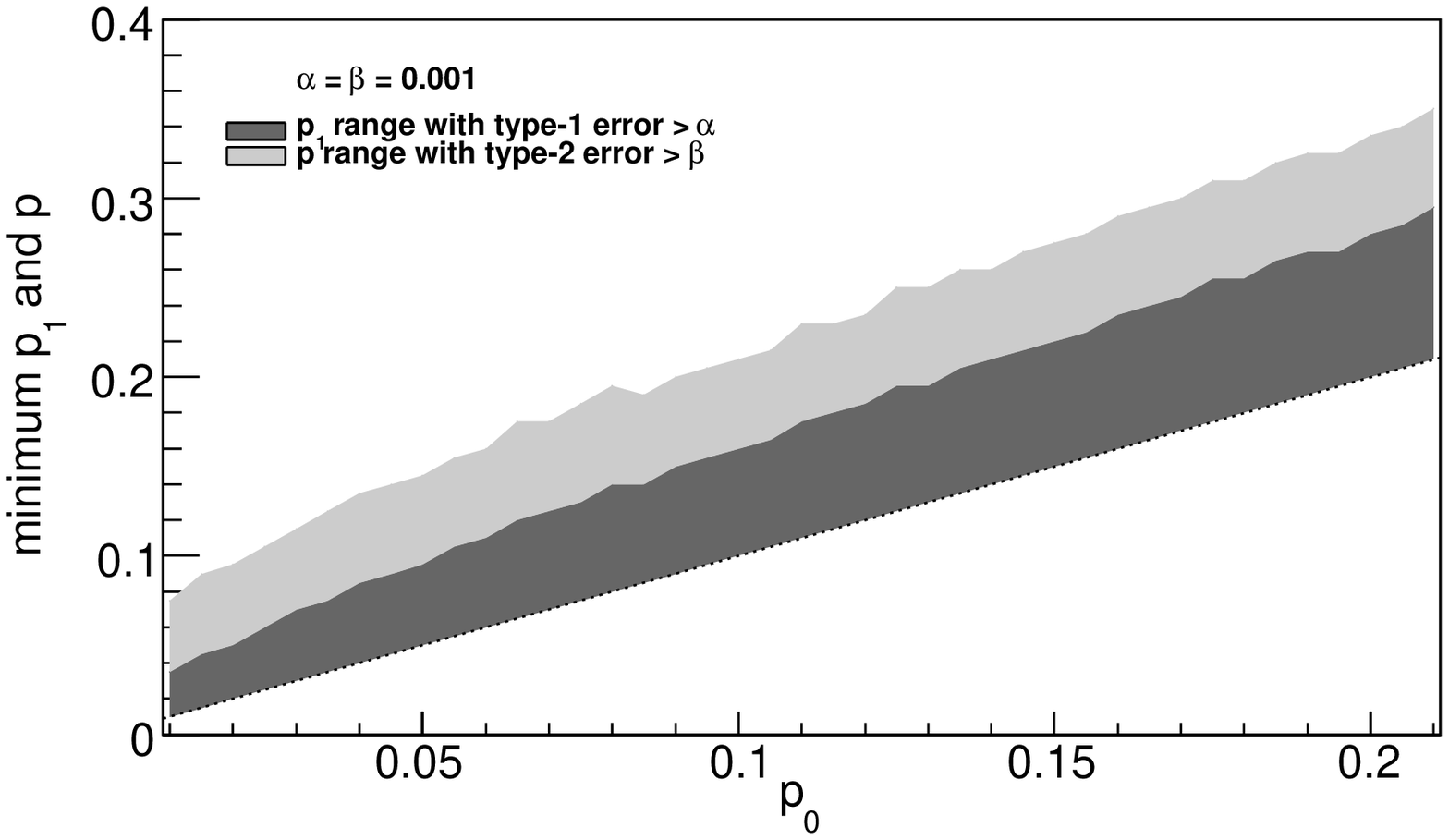}
\caption{Range for $p_1>p_0$ for which the ratio test terminates with 
type-1 error probabilities greater than $\alpha$ (dark grey), as a 
function of $p_0$.  Range for $p>p_1$ for which the ratio test terminates 
with type-2 error probabilities greater than $\beta$, as a function of $p_0$ (light 
grey).  The upper plot is for $\alpha=\beta=0.01$, the lower plot for 
$\alpha=\beta=0.001$.\label{fig:zones}}
\end{figure}

\textbf{Case 1: $\mathcal{H}_0$ is True:} First consider the case where the
null hypothesis is true, so that the correlation probability $p$ of the data is
equal to $p_0$.  The dark grey area in Fig.\,\ref{fig:zones} indicates, as a
function of $p_{0}$,  the range $p_1>p_0$ for which the ratio test terminates
with a type-1 error probability greater than $\alpha$.  Note that when
$p_1\simeq p_0$, there is a large fraction of data sets in which the test does
not come to a conclusion (rejection or acceptance of the null hypothesis) even
when the number of events $n$ exceeds 1000.  The fraction of undecided tests is
added to the type-1 error rate to give a conservative limit on $p_1$.
For all $p_1$ that fall above the dark grey area, the test terminates with a
type-1 error rate less than $\alpha$.  As expected, the dark grey range
is narrow, so the test is ``well-behaved'' if $p_1$ is chosen not too
close to $p_0$.  As an example, if the random correlation probability
$p_0=0.1$, then $p_1=0.14$ ($\delta=0.04$).  Any values for $p_1$ larger than
0.14 will of course also be well-behaved.

\textbf{Case 2: $\mathcal{H}_0$ is False:} We now consider the case where the
null hypothesis is false.  Choosing the values for $p_1$ determined with the
procedure outlined in ``Case 1,'' we use simulated data to find the minimum
signal probability $p_{\mbox{\scriptsize min}}$ for which the ratio test terminates with a
type-2 error probability less than $\beta$.  The light grey area in
Fig.\,\ref{fig:zones} depicts, as a function of $p_0$, the range of
$p_{\mbox{\scriptsize min}}>p_1$ for which the ratio test terminates with a type-2
probability greater than $\beta$.  For instance, when $p_0=0.1$ and
$\alpha=\beta=0.01$, for all signal probabilities $p>p_{\mbox{\scriptsize min}}=0.18$ the
ratio test will terminate with a type-2 error probability less than
$\beta$.  Note that the $p_{\mbox{\scriptsize min}}$ values given here are conservative,
since they not only require a type-2 error below $\beta$ in case of a
signal with strength $p_{\mbox{\scriptsize min}}$, but also a type-1 rate below
$\alpha$ \textit{and} a rejection or acceptance of $\mathcal{H}_0$ before the
sample size $n$ reaches 1000 when $\mathcal{H}_0$ is true.  This last
requirement slightly inflates the value of $p_{\mbox{\scriptsize min}}$.

The simulations of Cases 1 and 2 indicate that $p$ and $p_1$ must be larger
than $p_0$ if the test is to arrive at a decision in a reasonable amount of
time, and if the results are to be consistent with the error probabilities
$\alpha$ and $\beta$.  (To a much lesser extent, this second issue also exists
in Wald's original formulation of the ratio test, in which $p_1$ is treated as
a single alternative probability~\citep{Wald:1945,Wald:1947}.) Even so, the
amounts by which $p$ and $p_1$ should differ from $p_0$ are small enough that
they do not appreciably limit the usefulness of the method when a ``classical''
interpretation of $\alpha$ and $\beta$ is required.  We note that the existence
of small intervals above $p_0$ where such an interpretation is not possible are
a typical feature of sequential tests; see, for
example~\citep{Wald:1945,Wald:1947,Lewis:1994}.  It should be stressed,
however, that we have not demonstrated a circumstance where we are obtaining
some undesired values for $\alpha$ and $\beta$.  Rather, we have demonstrated
that marginalizing the likelihood is not the equivalent of inserting the right
value for $p$.

%In his original paper, Wald suggests a different approach for testing a point
%null hypothesis against a single-sided alternative.  Here, rather than
%marginalizing over the unknown correlation probability $p$, one chooses a
%single value $p_s>p_0$ and proceeds with the ratio test as if the point null
%hypothesis is tested against a {\it single} alternative probability $p_s$.  As
%in the method that uses a marginalized likelihood ratio, the fact that the data
%has a correlation probability $p$ that is in most cases not equal $p_s$ may
%result in type-1 error probabilities greater than $\alpha$ and type-2 error
%probabilities greater than $\beta$ for some $p_s$.  Again, $p_s$ can be chosen
%such that the probabilities do not exceed $\alpha$ and $\beta$.  This is
%typically the case if $p_s$ is chosen not too close to $p_0$.  

\subsection{Efficiency of the Ratio Test}

\begin{figure}
\plotone{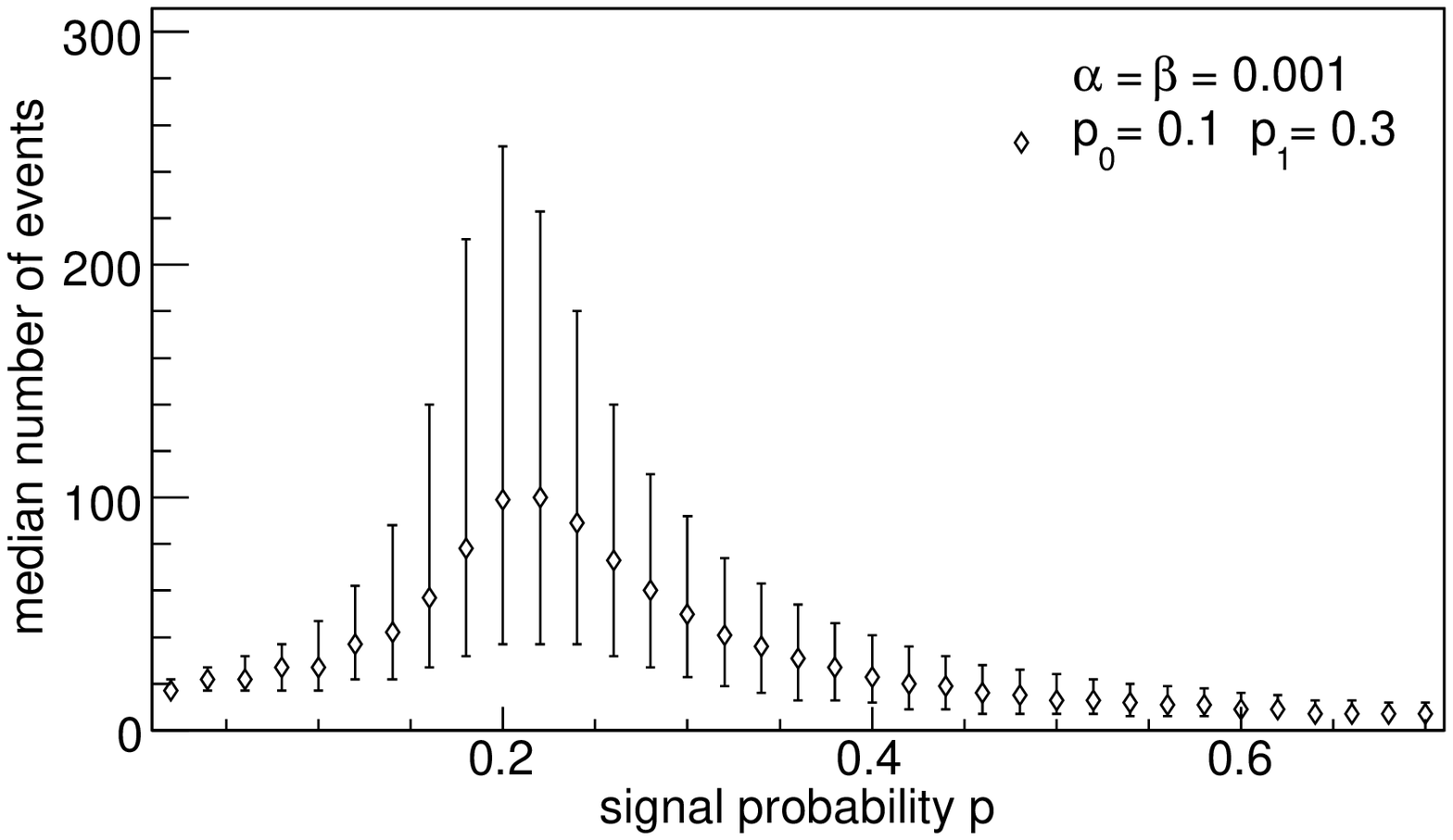}
\plotone{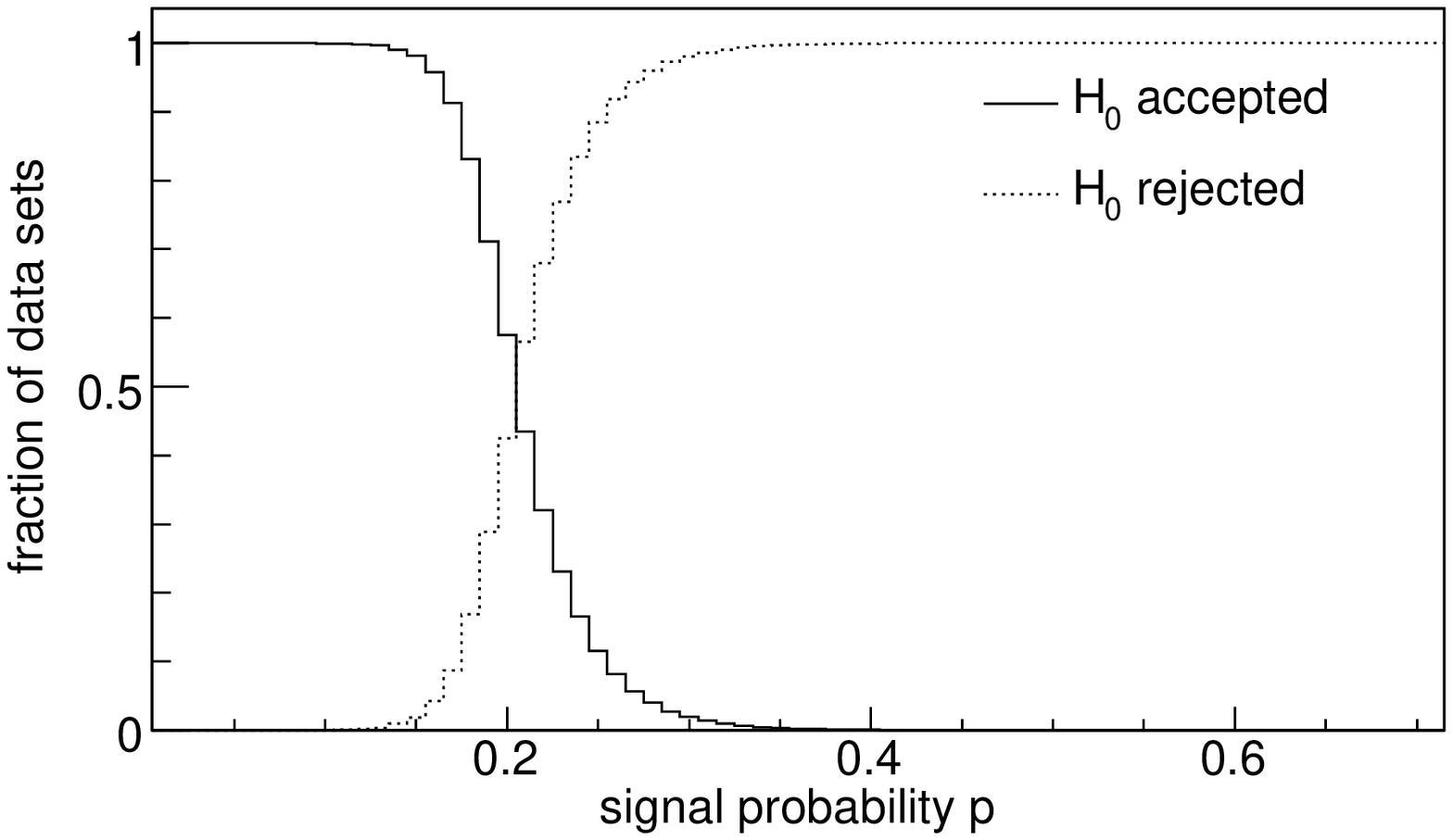}
\caption{{\it Top:}  Median number of events necessary for the sequential test 
to come to a conclusion, as a function of the signal probability $p$.  In this
example, the background probability is $p_0=0.1$, and $p_1=0.3$, $\alpha=\beta=0.001$.  
Error bars indicate the range that includes 68\,\% of the simulated data sets.  
{\it Bottom:}  For the same simulated data sets, fraction of data sets for which 
the null hypothesis is accepted (solid line) and rejected (dotted line) as a 
function of the signal probability $p$ for a background probability $p_0=0.1$. 
\label{fig:median}}
\end{figure}

An important aspect of a sequential test is its length, i.e., the number of
events $n$ necessary to reach a decision.  Fig.\,\ref{fig:median} shows an
example for the typical length of the test as a function of the signal
probability $p$.  In this example, the background probability is chosen as
$p_0=0.1$, the lower boundary of the marginalization is  $p_1=0.3$, and
$\alpha=\beta=0.001$.  For $10^5$ simulated data sets,
Fig.\,\ref{fig:median}\,(top) shows the median number of events required for a
termination of the test.  The error bars indicate the range that includes
68\,\% of the data sets.  In this example, the median size of a data set
required to accept the null hypothesis if it is true ($p_0=0.1$) is 27.  The
median size of a data set required to reject the null hypothesis if it is wrong
depends on $p$ and is large when $p$ is close to $p_0$.  Above $p\simeq 0.6$,
the median number reaches a plateau of about 7 events.

Fig.\,\ref{fig:median}\,(bottom) shows which decision is actually made,
depicting the fraction of data sets for which the null hypothesis ($p_0=0.1$)
is accepted and the fraction for which it is rejected, as a function of the
signal probability.  

\begin{figure}
\plotone{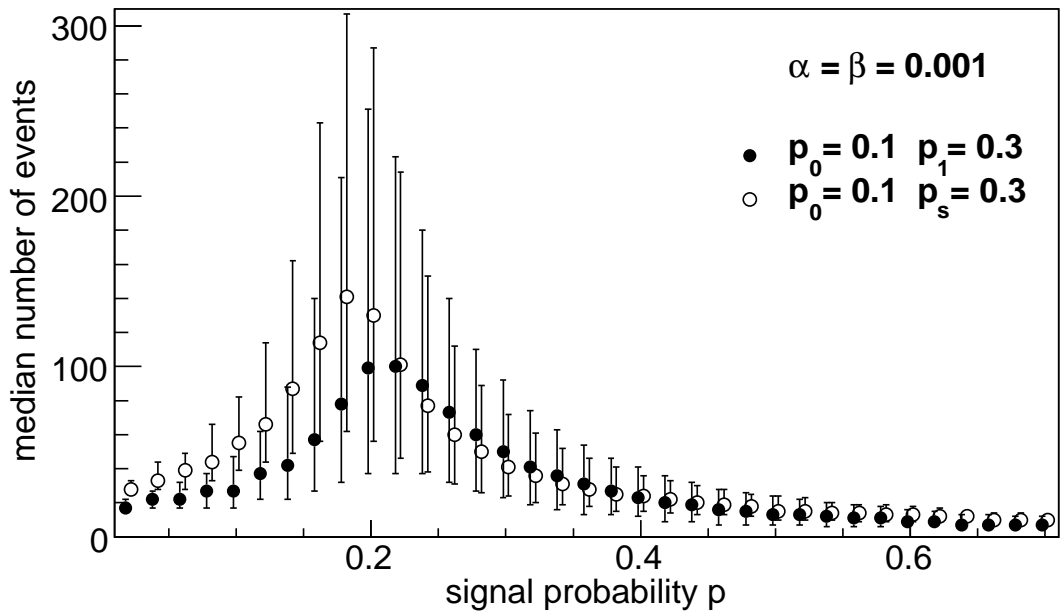}
\caption{Median number of events necessary for the Wald sequential test 
to come to a conclusion (open circles), as a function of the signal probability
$p$, compared to the marginalized likelihood ratio test (filled circles).  The
fixed point $p_1$ is the same in both cases.  For this example, the background
probability is $p_0=0.1$, and $p_1=0.3$, $\alpha=\beta=0.001$.  Error bars 
indicate the range that includes 68\,\% of the simulated data sets.  
\label{fig:median_wald}}
\end{figure}

Comparing the length of the test with the marginalized likelihood to Wald's
original test is not straightforward, since the length of each test depends on
the specifics of the problem, and because the probability $p_1$ has quite a
different meaning for the two methods.  However, we find that the marginalized
test tends to require fewer events when $p_1$ is the same in both tests.  For
the above example, the median number of events required to accept the null
hypothesis if it is true is 55 and thus twice as large as for the marginalized
likelihood ratio.  For signal probabilities $p>0.6$, the Wald test reaches a
plateau that is roughly comparable to the marginalized test.
Fig.\,\ref{fig:median_wald} shows the median number of events required for the
Wald test for $p_1=0.3$ and $\alpha=\beta=0.001$.

%Comparing the length of the test with the marginalized likelihood to Wald's
%original test which compares $p_0$ to a fixed signal probability $p_s$ is not
%straightforward, as the length of the test depends on our choice of $p_s$ just
%like the length of the marginalized test depends on our choice of $p_1$.  Both
%parameters can be chosen independently, and the Wald sequential test can end
%sooner or later than the marginalized test depending on what values are chosen.
%However, we find that the marginalized test tends to require fewer events if
%$p_1=p_s$.  For the above example, the median number of events required to
%accept the null hypothesis if it is true is 55 and thus twice as large as for
%the marginalized likelihood ratio.  For signal probabilities $p>0.6$, the Wald
%test reaches a plateau that is roughly comparable to the marginalized test.
%Fig.\,\ref{fig:median_wald} shows the median number of events required for the
%Wald test for $p_s=0.3$ and $\alpha=\beta=0.001$.

% -----------------------------------------------------------------------------
\section{Summary}\label{sec:summary}
% -----------------------------------------------------------------------------

We have outlined a sequential analysis technique for testing a point
null hypothesis with probability $p_0$ against a signal probability
$p$.  The method is based on the sequential analysis proposed 
in~\citet{Wald:1945,Wald:1947}, but replacing the likelihood ratio used
to evaluate the significance of a signal with one that marginalizes
the signal strength.

In many sequential tests, the signal strength is unknown when the test
starts.  Typically, the signal probability $p$ can in principle have any
value in the interval $\left[p_0,1\right]$.  Rather than choosing a fixed
threshold for $p$, as suggested in~\citet{Wald:1945,Wald:1947}, we have
argued that, in general, the better alternative is to marginalize $p$ and
account for our ignorance exactly.  In the marginalization of the signal
likelihood, the integration starts at some value $p_1=p_0+\delta$, where
$\delta$ is an \textsl{ad hoc} parameter reflecting the experimenter's belief
about the strength of the signal, the capability of his experiment,
and other \textsl{a priori} knowledge.

Because of the integration of the signal likelihood over a range in $p$, the
parameters $\alpha$ and $\beta$ have lost their intuitive meaning if the
method is applied to data sets where $p$ is fixed, as is typically the
case for real data.  However, we have shown that for most values of $\delta$
and $p$ that occur in correlation searches, the type-1 and type-2
error rates of the sequential analysis are consistent with the classical
interpretations of the probabilities $\alpha$ and $\beta$.

Note that we have run a test with one of two outcomes
(i.e., an acceptance or rejection of $\mathcal{H}_0$), defining $\alpha$ and 
$\beta$, rather than one outcome (say, only a rejection of $\mathcal{H}_0$) 
such as in~\citet{Darling:1968}.  The latter case supposes that we 
are only concerned about reporting a signal.  
However, it is important to state a null
result at some point in the interest of reducing reporting bias.
That is, it is important to ensure that 1\% of the 
results that claim an excess of events are indeed a 1\% effect. 

The sequential analysis technique proposed here is efficient, allows the
signal significance to be evaluated after the test has been fulfilled,
adheres to the likelihood principle, and rigorously accounts for our
ignorance of the signal strength.

\acknowledgments

We thank Diego Harari, Antoine Letessier-Selvon, and John A.J. Matthews for
valuable discussions and help.  This work is supported by the National Science
Foundation under contract numbers NSF-PHY-0500492 and NSF-PHY-0636875.

\appendix
\section{The Truncated Sequential Analysis Test}

In practice, the test must end.  It is supposed that a decision to 
accept or reject the null hypothesis must 
be made when $n=n_0$ if it has not been made already for $n\le n_0$.
Following the derivation of the modified errors for truncated tests 
in~\citet{Wald:1945}, $\alpha(n_0)$ and $\beta(n_0)$
are defined as the probabilities of errors of the first and second
kinds if the test is truncated at $n=n_0$.  The objective is then
to derive an upper bound on $\alpha(n_0)$ and $\beta(n_0)$ such 
that (1) the test ends prematurely and (2) 
$\mathcal{H}_1$ is accepted if $R_{n_0}>1$ and $\mathcal{H}_0$ is accepted in 
$R_{n_0}\le 1$.  In doing so, 
we find a suitable $\delta$ and $n_0$ where $\alpha$
and $\beta$ are small.

First, $\rho_0(n_0)$ is defined as the probability that, under the null hypothesis,
\begin{enumerate}
\item $B<R_{n_0-1}<A$
\item $1<R_{n_0}<A$
\item The sequential analysis would terminate with an acceptance
of $\mathcal{H}_0$ if allowed to continue.
\end{enumerate}
For the truncated test, we are rejecting the null hypothesis if
$1<R_{n_0}<A$.  In other words, $\rho_0(n_0)$ is the
probability of wrongly rejecting the null hypothesis 
when $1<R_{n_0}<A$ when it would have terminated with a 
rejection of the null hypothesis
wanted if we let the test continue.  This is 
added to the probability that the test would terminate wrongly if we let
it continue.  
Therefore, the upper bound on $\alpha(n_0)$ can be expressed as
\begin{eqnarray}
\alpha(n_0)\le \alpha + \rho_0(n_0).
\end{eqnarray}
Now if $\bar{\rho}_0(n_0)$ is 
simply the probability under the null hypothesis that $1<R_{n_0}<A$, 
then $\rho_0(n_0)<\bar{\rho}(n_0)$
and therefore 
\begin{eqnarray}
\alpha(n_0)\le \alpha + \bar{\rho}_0(n_0).
\end{eqnarray}
Similarly, $\rho_1(n_0)$ is defined as the probability that, 
under the ``signal'' hypothesis,
\begin{enumerate}
\item $B<R_{n_0-1}<A$
\item $B<R_{n_0}\le 1$
\item The sequential analysis would terminate with an acceptance
of $\mathcal{H}_1$ if allowed to continue.
\end{enumerate}
and
\begin{eqnarray}
\beta(n_0)\le \beta + \bar{\rho}_1(n_0).
\end{eqnarray}
where $\bar{\rho}_1(n_0)$ is defined to be 
the probability under the signal hypothesis that $B<R_{n_0}\le 1$.  

We then calculate $\bar{\rho}_0(n_0)$ explicitly.  
The probability of obtaining $R_{n_0}>1$ if the null hypothesis is true is
\begin{eqnarray}
\bar{\rho}_0(n_0)=
\sum_{k_{1+}}^{k_A} {n_0 \choose k} p_0^k(1-p_0)^{n_0-k}
\end{eqnarray}
where $k_{1+}$ is the minimum integer $k$ for which 
\begin{eqnarray}
\frac{\frac{1}{1-p_0-\delta}\int_{p_0+\delta}^1p^k(1-p)^{n_0-k}}{p_0^{k}(1-p_0)^{n_0-k}}>1
\end{eqnarray}
and $k_{A}$ is the maximum integer $k$ for which 
\begin{eqnarray}
\frac{\frac{1}{1-p_0-\delta}\int_{p_0+\delta}^1p^k(1-p)^{n_0-k}}{p_0^{k}(1-p_0)^{n_0-k}}<A
\end{eqnarray}

Similarly,
\begin{eqnarray}
\bar{\rho}_1(n_0)=
\frac{\sum_{k_B}^{k_{1-}} {n_0 \choose k} \frac{1}{1-p_0-\delta}\int_{p_0+\delta}^1p^{k}(1-p)^{n_0-k}}
{\sum_0^{n_0} {n_0 \choose k} \frac{1}{1-p_0-\delta}\int_{p_0+\delta}^1p^{k}(1-p)^{n_0-k}}
\end{eqnarray}
where $k_{1-}$ is the maximum integer $k$ for which 
\begin{eqnarray}
\frac{\frac{1}{1-p_0-\delta}\int_{p_0+\delta}^1p^k(1-p)^{n_0-k}}{p_0^{k}(1-p_0)^{n_0-k}}\le 1
\end{eqnarray}
and $k_B$ is the minimum integer $k$ for which 
\begin{eqnarray}
\frac{\frac{1}{1-p_0-\delta}\int_{p_0+\delta}^1p^k(1-p)^{n_0-k}}{p_0^{k}(1-p_0)^{n_0-k}}>B
\end{eqnarray}

\begin{figure}
\plotone{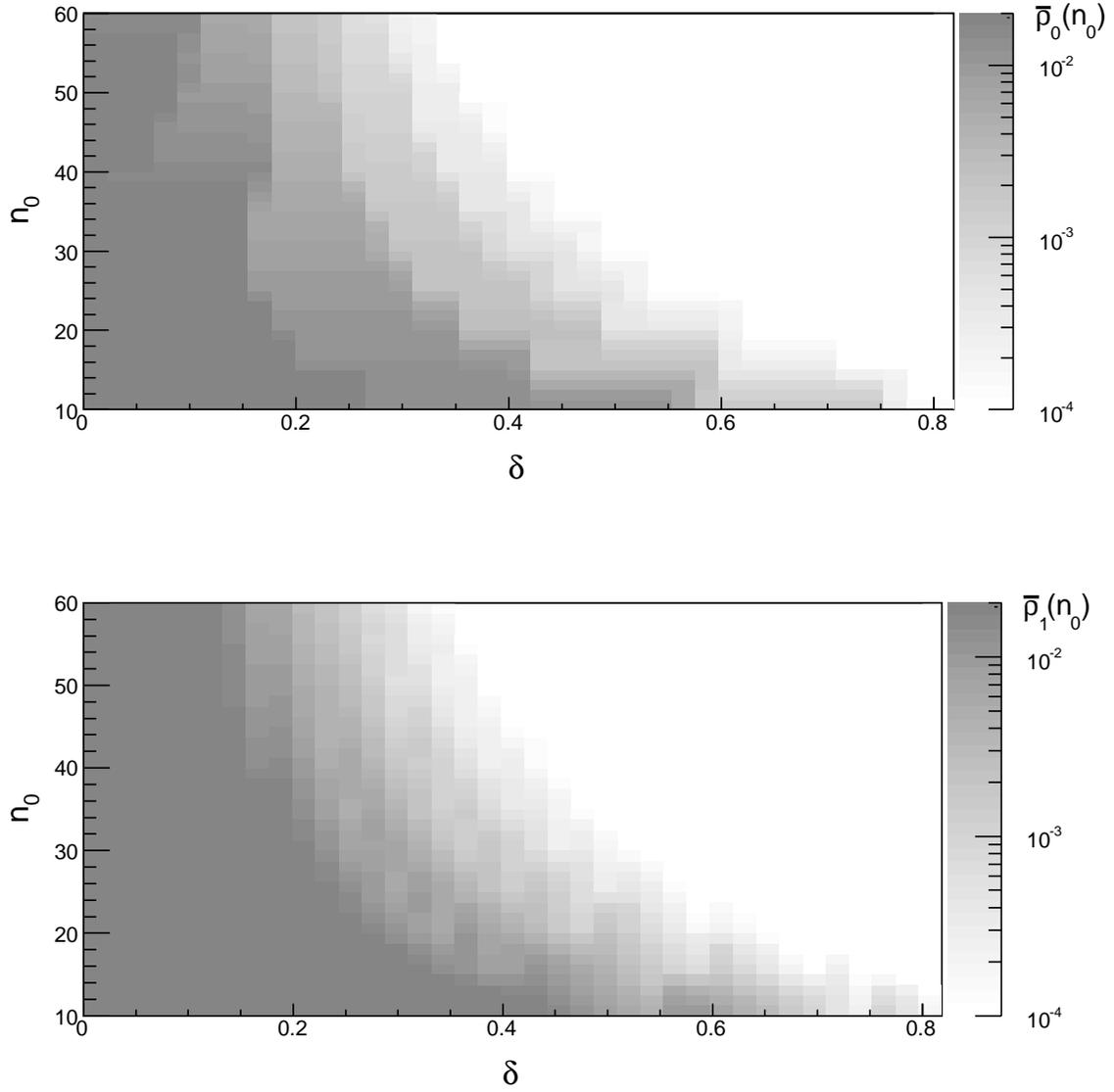}
\caption{The added error for $\alpha$, $\bar{\rho}_0(n_0)$, and $\beta$, $\bar{\rho}_1(n_0)$,
as a function of $\delta$, where $p$ is integrated from $p_0+\delta$ to 1, 
and the number of events at which the test is truncated, $n_0$.  }
\label{fig:rho}
\end{figure}

Under this scheme, Fig.\,\ref{fig:rho} shows $\bar{\rho}_0(n_0)$ 
and $\bar{\rho}_1(n_0)$ as a function of $\delta$ and $n_0$.
It shows that a rather large $\delta$ ($\sim 0.7$) is required to bring 
$\bar{\rho}_1(n_0)$ and $\bar{\rho}_1(n_0)$ to be less than 
$\alpha = \beta = 0.001$.  Further, if the calculation is
extended we find that
it would take $\sim 180$ events to bring 
$\bar{\rho}_1(n_0)$ and $\bar{\rho}_1(n_0)$ to be $\sim 0$
for any $\delta$.

\clearpage

\clearpage

\clearpage

\clearpage

\clearpage

\clearpage

\begin{thebibliography}{11}
\expandafter\ifx\csname natexlab\endcsname\relax\def\natexlab#1{#1}\fi

\bibitem[{Abbasi {et~al.}(2004)}]{Abbasi:2004ib}
Abbasi, R.~U. {et~al.} 2004, Astrophys. J., 610, L73

\bibitem[{Abbasi {et~al.}(2006)}]{Abbasi:2005qy}
---. 2006, Astrophys. J., 636, 680

\bibitem[{Anscombe (1954)}]{Anscombe:1954}
Anscombe, F.~J. 1954, Biometrics, 10, 89

\bibitem[{Armitage {et~al.}(1969)}]{Armitage:1969}
Armitage, P., McPherson, C.~K., \& Rowe, B.~C. 1969, J. Roy. Stat. Soc. A, 132, 235

\bibitem[{Berry (1987)}]{Berry:1987}
Berry, D.~A. 1987, Amer. Stat., 41, 117

\bibitem[{Darling {et~al.}(1968)}]{Darling:1968}
Darling, D.~A. \& Robbins, H. 1968, Proc. Nat. Acad. Sci. USA, 61, 804

\bibitem[{Gorbunov {et~al.}(2004)}]{Gorbunov:2004bs}
Gorbunov, D.~S., Tinyakov, P.~G., Tkachev, I.~I., \& Troitsky, S.~V. 2004, JETP
  Lett., 80, 145

\bibitem[{Jeffreys (1939)}]{Jeffreys:1939}
Jeffreys, H. 1939, Theory of Probability (London: Oxford University Press) 

\bibitem[{Lewis \& Berry (1994)}]{Lewis:1994}
Lewis, R.~J. \& Berry, D.~A. 1994, J. Amer. Stat. Assoc., 89, 1528

\bibitem[{Kass \& Raftery (1995)}]{Kass:1995}
Kass, R.~E. \& Raftery, A.~E. 1995, J. Amer. Stat. Assoc., 90 773

\bibitem[{Takeda {et~al.}(1999)}]{Takeda:1999sg}
Takeda, M. {et~al.} 1999, Astrophys. J., 522, 225

\bibitem[{Tinyakov \& Tkachev (2001)}]{Tinyakov:2001nr}
Tinyakov, P.~G. \& Tkachev, I.~I. 2001, JETP Lett., 74, 445

\bibitem[{Wald (1945)}]{Wald:1945}
Wald, A. 1945, Ann. Math. Stat., 16, 117

\bibitem[{Wald (1947)}]{Wald:1947}
---. 1947, Sequential Analysis (New York, NY: John Wiley and Sons)

\end{thebibliography}
\end{document}